\documentclass[twocolumn,showpacs,preprintnumbers,amsmath,amssymb,footinbib]{revtex4-1}
\usepackage{graphicx}
\usepackage{subfig}
\usepackage{color}

\begin{document}

\title{Type I parametric down conversion of highly focused Gaussian beams in finite length crystals}

\author{Yasser Jeronimo-Moreno}
\author{R. J\'auregui}
\affiliation{Instituto de F\'isica, Universidad Nacional Aut\'onoma de M\'exico, Apartado Postal 20-364, M\'exico DF 01000, M\'exico\\$^*$Corresponding author: yjeronimo@fisica.unam.mx, rocio@fisica.unam.mx}

\begin{abstract}
We study the  correlations in wave vector space of photon pairs generated by  type I  spontaneous parametric down conversion  using a Gaussian pump beam. We analyze both moderate focused and highly focused regimes taking special attention to the angular spectrum  and the conditional angular spectrum. Simple analytic expressions are derived  that allow us to study in detail the dependence of these spectra on the waist of the source and the length of the nonlinear crystal. These expressions are in good agreement with numerical expectations and reported experimental results. They are used to make a systematic search of optimization parameters that improve the feasibility of using highly focused Gaussian beams to generate idler and signal photons with predetermined mean values and spread of their transverse wave vectors.
\end{abstract}

\maketitle

\section{Introduction}

The process of spontaneous parametric down conversion (SPDC) \cite{Burnham1970}, where a pump beam interacts in a non linear crystal and individual photons of the pump decay in a pair of photons, is one of the most reliable sources of photons with predetermined properties\cite{monken2010}.  In particular, the two photon state may exhibit spatial entanglement   \cite{Rubin1996,Kwiat1997,Neves2009} that can be manipulated via the modification of the spatial structure of the pump beam  and/or the crystal length. Varying them, it is even possible to optimize the coupling efficiency  to other  optical elements like optical fibers\cite{Bovino2003, dragan2004,Castelletto2005,Andrews2004,Ljunggren2005,Kolenderski2009, minozzi2013}.

Expressions for the main properties of the emission cone and their dependence on the crystal length have been the subject of intense research  because of their potential use in the quantum engineering of SPDC \cite{Rubin1996,uren2003,Lee2005,Bennink2006,Suzer2008,Grice2011,Pires2011,uren2013}. In this work, we study the  spatial properties of the emitted photons via the angular spectrum (AS) and the conditional angular spectrum (CAS). Using the exact expressions for the birefringent dispersion relations, we numerically evaluate these spectra  both in the moderate focused and highly focused regimes of a Gaussian pump beam. When reliable  approximations for the phase matching function are used, we also obtain simple analytical expressions that
reproduce accurately the numerical expectations and reported experimental results; they allow a better understanding of the effects of the pump beam waist and crystal length in both the AS  and CAS functions. We also study the usage of these expressions for the optimization of the SPDC process in the generation of idler and signal photons localized in predetermined regions of the wave vector space.

\section{Theory.}

To first order in perturbation theory, the SPDC state  is given by \cite{monken2010}:
\begin{equation}
\label{E:St}
\vert \Psi\rangle=  \vert \mbox{vac}\rangle
+\int d\omega_s d\omega_id^2\mathbf{k}_{\bot,s} d^2\mathbf{k}_{\bot,i}
F_{s,i}\vert \alpha_p; 1_s;1_i\rangle,
\end{equation}
where  $F_{s,i}=F(\mathbf{k}_{\bot,s},\omega_s;\mathbf{k}_{\bot,i},\omega_i)$ represents the joint amplitude  expressed as  a function of the frequency $\omega_b$ and the transversal wave vectors $\mathbf{k}_{\bot,b}$ for the signal and idler modes ($b=s,i$). It has the following
structure:
\begin{equation}
F(\mathbf{k}_{\bot,s},\omega_s;\mathbf{k}_{\bot,i},\omega_i)=g\mathfrak{E}(\mathbf{k}_{\bot,p},\omega_p)  f (\mathbf{k}_{\bot,s},\omega_s;\mathbf{k}_{\bot,i},\omega_i)\label{E:F}
\end{equation}
with $g$  the product of the effective second order susceptibility and the normalization factors of the quantized pump, signal and idler  photon states. $\mathfrak{E}$ is the amplitude function of a Gaussian coherent pump beam,
\begin{eqnarray}
\label{eq:Gauss}
{\mathfrak{E}}(\mbox{k}_{x,p} ,\mbox{k}_{y,p},\omega_p)&=&
 \alpha(\omega_p)\cdot\mbox{e}^{-(k_{x,p}W_x/2)^2(1-i\frac{z_o}{z_{Rx}})}\nonumber\\
 &\times&\mbox{e}^{-(k_{y,p}W_y/2)^2(1-i\frac{z_o}{z_{Ry}})}
\end{eqnarray}
with  $\alpha(\omega_p)$ its spectral amplitude, $W_x$ and $W_y$ the values of waist pump along $x$-axis and $y$-axis respectively,
$z_o$ determines the focal plane and $z_{R}$ are the Rayleigh lengths along $x$-axis and $y$-axis. Energy conservation implies $\omega_p=\omega_s+\omega_i$. For a wide crystal, the
conservation of transversal momenta yields the condition $\mathbf{k}_{\bot,p}=\mathbf{k}_{\bot,i}+\mathbf{k}_{\bot,s}$.
In Eq.~(\ref{E:F}), $f$  corresponds to the phase matching function:
\begin{equation}
f (\mathbf{k}_{\bot,s},\omega_s;\mathbf{k}_{\bot,i},\omega_i)=  L\mbox{sinc}\left(L \Delta \mbox{k}_z/2\right)\mbox{e}^{iL\Delta \mbox{k}_z /2},
\end{equation}
where $L$ is the crystal length and the mismatch term is $\Delta k_z=k_{z,p}-k_{z,s}-k_{z,i}$.

In this paper we concentrate  on  type-I SPDC and degenerated emission, i.e., $\omega_p= \omega=2\omega_s=2\omega_i$. Thus, the emitted photons  have ordinary polarization and  their dispersion relation  is
\begin{equation}
\mbox{k}_{z,b}^O=\sqrt{\epsilon_\bot  \omega^2/4 c^2 - k_{\bot,b}^2}\quad\quad b=i,s
\label{eq:dispord}
\end{equation}
with $c$  the velocity of light in vacuum, $\epsilon_\bot$ is  the permeability function transversal to the ordinary plane defined by the optical axis $\mathbf{a}=(0,\mbox{a}_y,\rm{a}_z)$;  the ordinary refraction index is $n_o = \sqrt{\epsilon_\bot}$. The permeability coefficient parallel to the optical
axis  $\epsilon_\parallel$ yields  the  dispersion relation for the pump wave which evolves in the extraordinary plane:
\begin{eqnarray}
\mbox{k}_z^{\tiny{E}}(\mathbf{k}_\bot,\omega)&=& -\beta \mathbf{a}\cdot\mathbf{k}_\bot  +\frac{\omega}{c}\mbox{n}_{eff}\sqrt{1
-\frac{k_\bot^2 c^2}{\omega^2}\eta},\label{eq:dispextraord}\\
\mbox{n}_{eff} &=&  \sqrt{\frac{\epsilon_\bot\epsilon_\parallel}{\epsilon_\bot + \Delta\epsilon{\rm a}_z^2}},\quad
 \beta =\frac{\Delta\epsilon {\rm a}_z}{\epsilon_\bot + \Delta\epsilon{\rm a}_z^2},\\
 \eta &=& \frac{1}{\epsilon_\bot + \Delta\epsilon{\rm a}_z^2}, 
\end{eqnarray}
$\Delta\epsilon= \epsilon_\parallel - \epsilon_\bot$. The $\beta$ term is responsible for the deviation of the Poynting vector with respect to the pump wavefront inside the crystal, the so called walk off effect\cite{Walborn2004,torres,monken2008,Walmsley2010}. The $\eta$ term gives rise to  astigmatic effects \cite{monken2008}. In the limit of normal incidence, this equation reduces to the  expression of the effective refractive index experienced by a paraxial pumping extraordinary wave,
$
\mbox{k}_z^{\tiny{E}}({\mathbf{k}}_\bot \sim \mathbf{0}) = \mbox{n}_{eff}\omega/c.
$
The permeability coefficients $\epsilon_\parallel$ and $\epsilon_\bot$ depend on the light frequency and some experimental set ups are implemented assuming ${n}_{eff}(\lambda) = n_o(2\lambda)$. We shall consider the general case where this equation is not taken for guaranteed.

The distribution of the signal/idler photons in the wave vector domain  is the angular spectrum (AS):
\begin{equation}
\label{E:Rs}
R_s(\mathbf{k}_{s},\omega_s; \omega_i)= \int d \mathbf{k}_{\bot,i}  |F(\mathbf{k}_{\bot,s},\omega_s ;\mathbf{k}_{\bot,i},\omega_i)|^2
\end{equation}

The conditional angular spectrum (CAS), which is a function of $\mathbf{k}_{\bot,s}$ and $\mathbf{k}_{\bot,i}$, is defined as:
\begin{equation}
\label{E:Rc}
R_c (\mathbf{k}_{\bot,s},\omega_s ;\mathbf{k}_{\bot,i}, \omega_i)= |F(\mathbf{k}_{\bot,s},\omega_s ;\mathbf{k}_{\bot,i},\omega_i)|^2
\end{equation}
This function represents the probability to detect a signal photon with wave vector ${\mathbf{k}}_{\bot,s}$
in coincidence with an idler photon with wave vector ${\mathbf{k}}_{\bot,i}$. In the paraxial regime ($W  k_{z,p}  \gg 1$)
the pump  beam  can be approximated by a plane wave, and, if we consider a long crystal, the SPDC process results in a
strict momentum conservation that correlates the observation of an individual signal photon with $\mathbf{k}$-vector
to a single idler photon with a well defined {\bf k}-vector. In experimental situations involving the small but finite
transverse dimensions of the pump beam and a usually wide but not so long crystal, there is a set of relevant pump wave
vectors $\{\mathbf{ k}_p\}$ that are close to satisfy the phase matching condition $\Delta{\bf k}_z =0$ for given idler and signal wave vectors ${\bf k}_{i,s}$.
Notice that $R_c$ results independent of the phase factor which contains the information about the Rayleigh lengths. That is,
$R_c$ depends just on the amplitude of the Fourier content of the pump beam.

In order to obtain approximate  expressions for the AS and CAS  functions, we   make a first order Taylor description of the phase mismatch,
\begin{eqnarray}
\Delta k_z&\sim & \kappa - \mathbf{d} \cdot (\mathbf{k}_{\bot,s} +\mathbf{k}_{\bot,i}),\\
\kappa &=&(\omega/c) (n_{eff} -n_o) + (2 c/n_{o}\omega) k_{\bot,s}^2, \\
\mathbf{d} &=& \beta \mathbf{a} + (2c/n_o \omega) \mathbf{k}_{\bot,s}.
\label{E:delta}
\end{eqnarray}
Note that in the absence of walk off effects $\mathbf{d}$ would point in the $\mathbf{k}_{\bot,s}$ direction.
Approximating the function $\mbox{sinc}(x)$ by a Gaussian function $\mbox{exp}[-(\gamma x)^2]$, $\gamma=0.4393$, the  expression for the CAS becomes:
\begin{eqnarray}
&&R_c (\mathbf{k}_{\bot,s},\omega_s;\mathbf{k}_{\bot,i}, \omega_i)= \vert  gL\alpha(\omega_s +\omega_i)\vert^2\times \nonumber\\&&\mbox{e}^{- \sigma_x^2 (k_{x,s}+k_{x,i})^2 -  \sigma_y^2 (k_{y,s} + k_{y,i})^2}
\mbox{e}^{-\gamma^2 L^2(\kappa^2 -2\kappa \mathbf{d}\cdot (\mathbf{k}_{\bot,s}+\mathbf{k}_{\bot,i} ))/2}
\nonumber\\&&\mbox{e}^{- (\gamma^2 L^2/2) (2 d_x d_y (k_{x,s}+k_{x,i}) (k_{y,s} + k_{y,i}))}
\label{E:F2}
\end{eqnarray}
$
\sigma_x^2 =(W_x^2 + \gamma^2 L^ 2 d_x^2)/2 , \quad\quad \sigma_y^2 =(W_y^2 + \gamma^2 L^ 2 d_y^2)/2.$

   The effective width of the Gaussian factor that modulates the CAS function has two contributions, one due solely to  the pump geometry and another highly dependent on the crystal optics and geometry. The latter includes a term $\sim L(\Delta\epsilon {\rm a}_z\mathbf{a}/(\epsilon_\bot + \Delta\epsilon{\rm a}_z^2))$. In the paraxial regime the   CAS function will be determined by the geometry of the Gaussian pump beam whenever $W_{b} \gg L(\Delta\epsilon {\rm a}_z{\rm a}_b/(\epsilon_\bot + \Delta\epsilon{\rm a}_z^2))$. In fact, for long crystals, second order terms proportional to  $k^2_{\bot,p}$
in the extraordinary dispersion relation, Eq.~(\ref{eq:dispextraord}), could become relevant. In such a case a third term contributes to the effective width of the Gaussian pump beam:  $$\tilde W_{b}^2 = W_{b}^2 +\frac{\gamma^2 L^2 n_o^2}{\epsilon_\bot + \Delta\epsilon\mathrm{a}_z^2}\left(\frac{n_{eff}}{n_o}-\frac{n_{eff}^2}{n_o^2}\right), \quad b=x,y,$$  replaces
$W_{b}^2$ in Eq.~(14).
 In general,
for a symmetric pump beam $W_x=W_y=W$, the CAS function may exhibit an asymmetric profile due to the orientation of the birefringent axis as encoded in $\mathbf{d}$.

We shall now illustrate  the CAS function by (i) choosing the value of the transversal wave vector of the idler photon $\mathbf{k}_{\bot,i}^0$ that maximizes the counts and (ii) reporting the corresponding distribution of the signal transverse wave vector, $\mathbf{k}_{\bot,s}$.
We take parameters from particular reported experimental situations\cite{uren2013}. This  allows to confirm the
reliability of our numerical simulations. Thus,  we consider a $1$ mm length BBO crystal, cut for type-I phase matching for  degenerated emission with angle $\theta_\mathbf{a}=29.3^\circ$, (the optical axis is defined by $\mathbf{a}= (0,\sin\theta_\mathbf{a},\cos\theta_\mathbf{a})$) ;  the quasi monochromatic pump beam has a wavelength centered at $406.99$ nm. Calculations are performed both using the exact expressions for the birefringent dispersion relations for the ordinary  wave, Eq.(\ref{eq:dispord}), and extraordinary wave, Eq.(\ref{eq:dispextraord}), as well as using the approximate analytical equation (\ref{E:F2}).

\begin{figure*}
\centering
\subfloat[]{\label{F:CAS185}\includegraphics[width=0.3\textwidth]{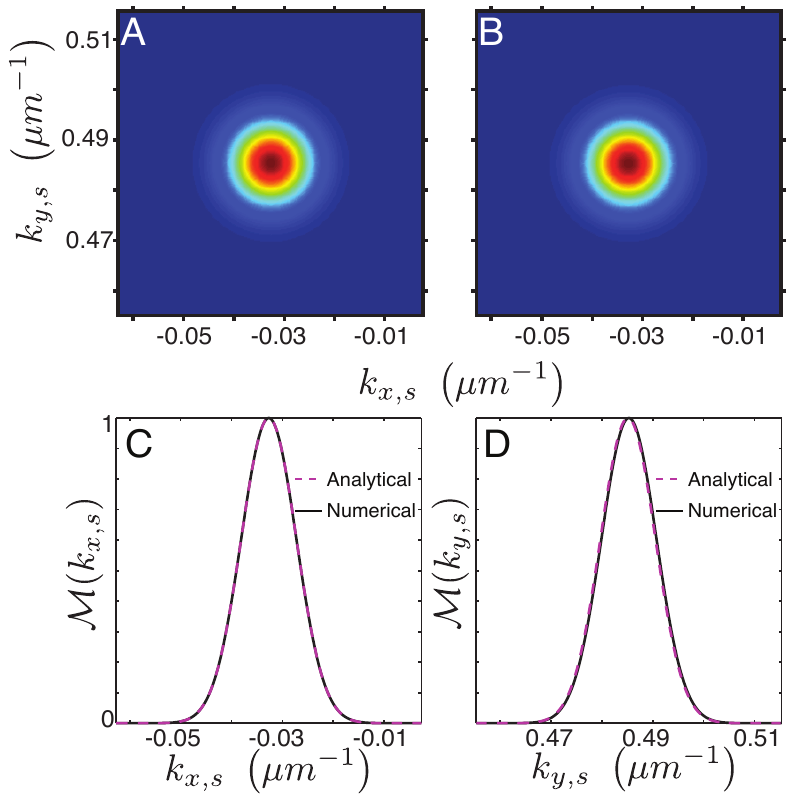}}
\subfloat[]{\label{F:CAS35}\includegraphics[width=0.3\textwidth]{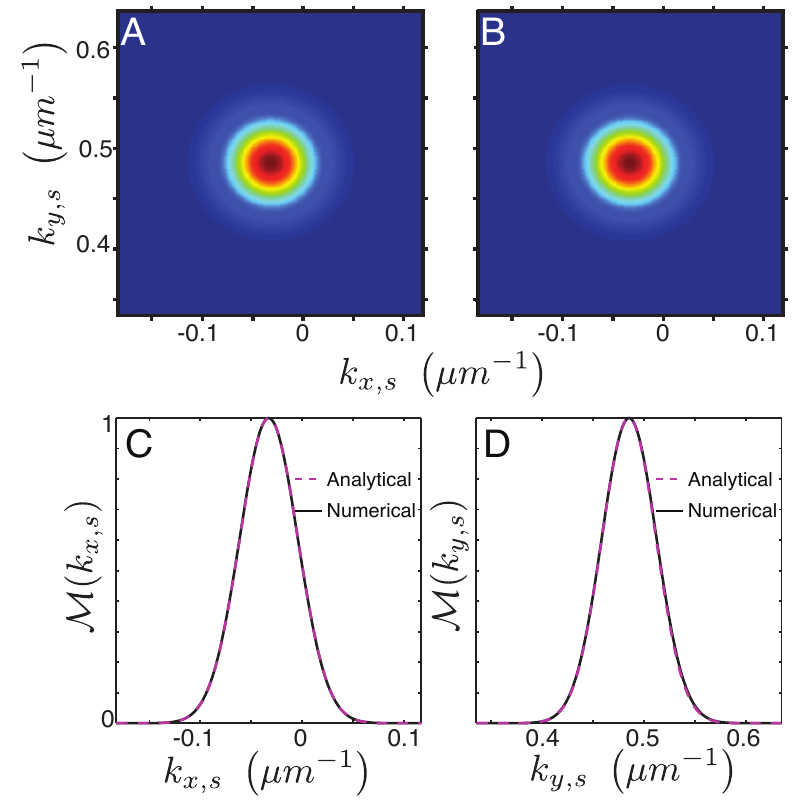}}
\subfloat[]{\label{F:CAS5}\includegraphics[width=0.3\textwidth]{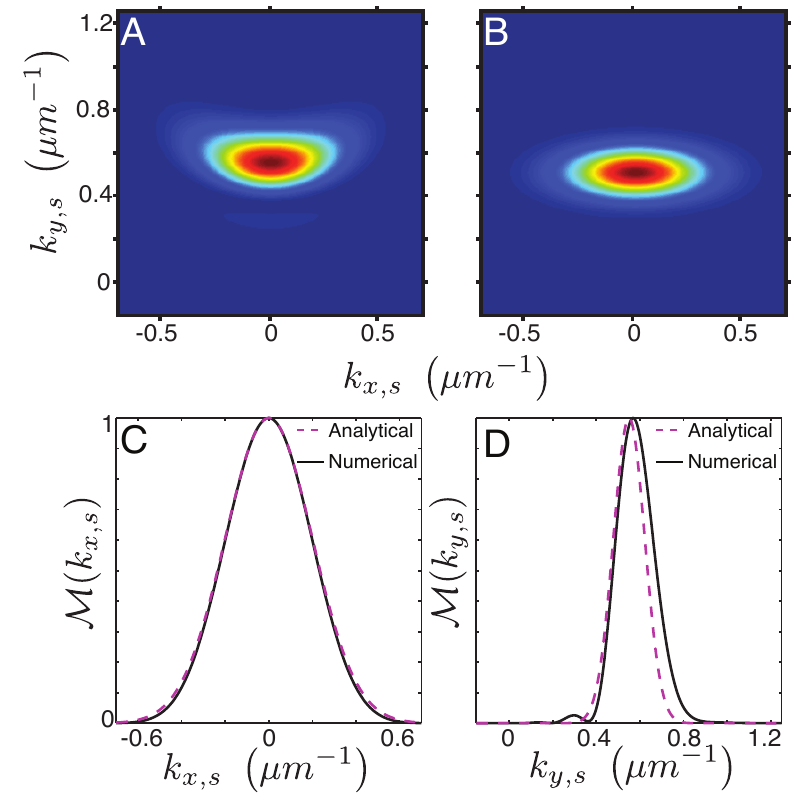}}
\caption{ Conditional angular spectra (CAS) for a $1$ mm BBO crystal using a symmetric Gaussian pump with waist (a)$W=185\mu$m; (b)$W=35\mu$m;(c)$W=5\mu$m. The transverse wave vectors of the corresponding idler photon are (a) $k_{x,i} = 0.033\mu$m$^{-1}$, $k_{y,i}=-0.485 \mu$m$^{-1}$ (b) $k_{x,i} = 0.033\mu$m$^{-1}$, $k_{y,i}=-0.485 \mu$m$^{-1}$ and (c) $k_{x,i} = 0.000\mu$m$^{-1}$,
 $k_{y,i}=-0.506 \mu$m$^{-1}$.  These wave vectors were chosen so that the counts  are maximized. In (A) CAS is calculated numerically, and in (B) analytically. Figs. (C)-(D) are the marginal distributions calculated numerically (solid line) and  analytically  (dashed line). }\label{fig:1}
\end{figure*}

Fig.\ref{F:CAS185}(A) shows the numerical CAS function for a symmetric pump waist $W_x$=$W_y$=$W$=$185\mu$m. As expected from Eq.~(\ref{E:F2}), for a pump waist that validates the paraxial regime and $W\gg\gamma L \vert\mathbf{d}\vert$, the CAS describes highly localized momentum  correlations determined mainly by the pump waist $W$. Fig.\ref{F:CAS185}(B) illustrates the CAS function calculated from the approximate Eq.(\ref{E:F2}). In order to make  a more clear comparison between numeric and analytic calculations, Figs.\ref{F:CAS185}(C)-(D) show the marginal distributions
$
\mathcal{M}(k_{b,s})=\int dk_{j\ne b,s} R_c(\mathbf{k}_{\bot,s},\omega_s;\mathbf{k}_{\bot,i}^0,\omega_i^0)
$ for $b,j=x,y$. The resulting agreement between the numerical and analytic calculations is also very good.

In Fig.\ref{F:CAS35}(A) the CAS is illustrated for a symmetric focused pump beam $W=35 \mu$m. Notice that the spread of the marginal distributions is, in this case, about twice that observed for a  pump beam with $W=185 \mu$m. This is a direct consequence of the fact that a higher focusing necessarily involves the incorporation of a wider distribution of pump wave vectors.

For a highly focused beam, as that shown in Fig.~\ref{F:CAS5} where $W=5\mu$m, the CAS function profile spreads even more and the conditional angular spectra becomes more sensitive to  the crystal  length and optical parameters  as described by the analytical expression, Eq.~(\ref{E:F2}),  Notice that although this approximate distribution does not reproduce in all details its numeric analog, it has the correct central point and width. The discrepancy is mainly due to the usage of the approximate dispersion relation as described above.  Another effect of the crystal is present in the small oscillations in the marginal distribution $\mathcal{M}(k_{y,s})$ in Fig.\ref{F:CAS5}(D) that were lost in the replacement of the phase matching sinc by a Gauss function. 
Besides, for this highly focused pump beam there is an observable difference between the absolute mean values of the y-component of the wave vector of the idler  photon $k_{y,i}=-0.51\mu$m$^{-1}$ and of the signal  photon $\langle k_{y,s}\rangle \sim 0.57\mu$m$^{-1}$.
The CAS for $W=185\mu$m and $W=35\mu$m  can be successfully compared with Fig.~2 of Ref.~ \cite{uren2013}.

\begin{figure*}
\centering
\includegraphics[width=0.55\textwidth]{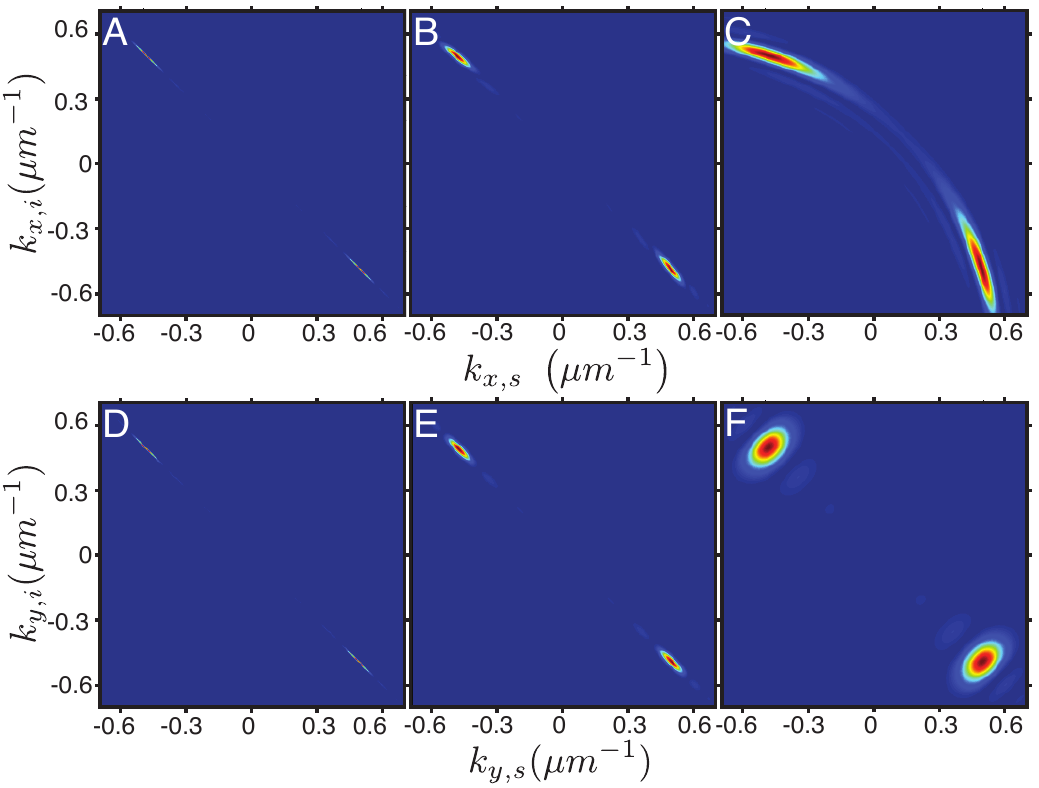}
\caption{ In the first row the conditional spectrum that arise from demanding $k_{y,s}=k_{y,i}=0$ is illustrated for pump waists
(a)$W=185\mu$m; (b)$W=35\mu$m;(c)$W=5\mu$m. In the second row, the conditional spectrum that arises from demanding $k_{x,s}=k_{x,i}=0$ is illustrated for the same pump waists. The general parameters are the same as those reported in Fig.~\ref{fig:1}.}
\label{fig:cross}
\end{figure*}

 In order to illustrate other interesting features of the CAS, we now consider the particular case in which the condition $k_{x,s}=k_{x,i}=0$ is imposed in the idler and signal photons and compare it to the case  $k_{y,s}=k_{y,i}=0$. This results in a conditional distribution for $k_{y,s}$ and $k_{y,i}$, and  a conditional distribution for $k_{x,s}$ and $k_{x,i}$ respectively. The CAS's are illustrated in Fig.~\ref{fig:cross} taking similar parameters as those used in previous examples. For $W=185\mu$m the signal and idler transverse wave vectors are restricted to almost opposite directions, so that for
$k_{x,s}=k_{x,i}=0$, $k_{y,s} \sim -k_{y,i}$ with $\vert k_{y,i}\vert\sim 0.5\mu$m$^{-1}$. As the pump beam waist decreases,
the anisotropic role of higher values of $\mathbf{k}_{\bot, p}$ in the phase matching condition is more relevant. These has two consequences. One of them is the increasing width on the CAS distributions. The other is that such distributions become clearly different for the condition $k_{x,s}=k_{x,i}=0$ and the condition $k_{y,s}=k_{y,i}=0$.

An approximate closed expression for the angular spectra distribution can also be obtained.  For simplicity we  consider a symmetric  pump profile ($W_x=W_y=W$). Using the  conservation of transverse momentum, writing the pump integration variable $\mathbf{k}_{\bot,p}$ in polar coordinates, and performing a rotation of the integration variable  by an angle $ \theta_r = {\rm arc} \cos(d_x/\vert\mathbf{d}\vert)$, one obtains:
\begin{equation}R_s(\mathbf{k}_{s})=\vert g L \alpha \vert^2\mbox{e}^{
 -\sigma_{\rm{AS}}^{-2} \left(k_{\bot,s}^{2} -
r_{\rm{AS}}^2\right)^2  }  \int_{0}^{2\pi}d\theta \int_0^{\tilde{k}_\bot} dk_\bot k_\bot \mbox{e}^{-u}
\end{equation}\begin{eqnarray}
r_{\rm{AS}}^2 = \frac{1}{2}\left(\frac{ n_o\omega}{c} \right)^2\left(1-\frac{n_{eff}}{n_o}\right)&,&
\sigma_{\rm{AS}}^{-2}=  \frac{2(\gamma Lc/n_o\omega)^2}{1 + \left(\gamma L\vert{\mathbf{d}}\vert / W\right)^2}  \nonumber \\
\tilde{k}_\bot ={\frac{\omega \sqrt{\epsilon_\bot\epsilon_\parallel}}{c\sqrt{\epsilon_\bot +\Delta\epsilon {\rm a}_\bot^2\cos^2\theta}}} &,&\zeta_x= \frac{\kappa\vert{\mathbf{d}}\vert}{\vert \mathbf{d} \vert^2+ (W/\gamma L)^2}\label{eq:rassigma}\end{eqnarray}
$u=((W^2+(\gamma L\vert{\mathbf{d}}\vert)^2) \left(k_\bot \cos \theta -\zeta_x \right)^2+ W^2k_\bot^2\sin^2 \theta)/2$.
By evaluating $\partial u(k_\bot,\theta)/\partial k_\bot$ it can be  shown that
\begin{eqnarray}
R_s(\mathbf{k_{s0}})=\vert g L \alpha(\omega_p) \vert^2{\rm e}^{- \sigma_{\rm{AS}}^{-2} \left(k_{\bot,s}^{2} -
r_{\rm{AS}}^2\right)^2}&&
\Big[\int_0^{2 \pi} d\theta \frac{1-\mbox{e}^{-u}}{\mbox{den}(\theta)} 
\nonumber\\+ W\zeta_x \int_0^{2 \pi} \frac{d\theta \cos \theta}{\mbox{den}(\theta)} \int_0^{^{\tilde{k}_\bot}}dk_\bot \mbox{e}^{-u}\Big]&,&\label{eq:ASAPPROX}
\end{eqnarray}
$\mbox{den}(\theta) =(W^2+(\gamma L\vert{\mathbf{d}}\vert)^2) \cos^2 \theta + W^2 \sin^2 \theta$.
The second integral over the variable $dk_\bot$ can be expressed in terms of error functions, and the remaining integral over the variable  $\theta$ could be evaluated numerically. Alternatively we can also make an approximate evaluation of Eq.~(\ref{eq:ASAPPROX}). At the  angle
$\theta_{est}={\rm arc} {\rm cos}(\kappa/\vert\mathbf{d}\vert k_{\bot,s}c)$ the exponent $u(k_\bot,\theta)$ takes an stationary value: $\partial u(k_\bot,\theta_{est})/\partial\theta = 0$. We replace
$u(k_\bot,\theta)$ by $u(k_\bot,\theta_{est}) =  \frac{W^2}{2}k_\bot^2 - \frac{\gamma^2 L^2\kappa ^2}{2}\Big[1-\frac{1} {1 +(W/\gamma L\vert{\mathbf{d}}\vert)^2}\Big]$
in Eq.~(\ref{eq:ASAPPROX}). Performing the remaining integrals we obtain:
\begin{equation}
R_s(\mathbf{k_{s0}}) \approx \frac{\pi\vert g L \alpha(\omega_p) \vert^2 {\rm e}^{- \sigma_{\rm{AS}}^{-2} \left(k_{\bot,s}^{2} -
r_{\rm{AS}}^2\right)^2 }}{W^2\sqrt{1 + \left(\gamma L\vert \mathbf{d} \vert / W\right)^2}} \Big(1 - e^{-u(\tilde{k}_\bot,\theta_{est})}\Big).
\label{E:RsAn}
\end{equation}
In most practical situations
 $1\gg e^{-u(\tilde{k}_\bot,\theta_{est})}$.
\begin{figure*}
\centering
\subfloat[]{\label{F:AS185}\includegraphics[width=0.3\textwidth]{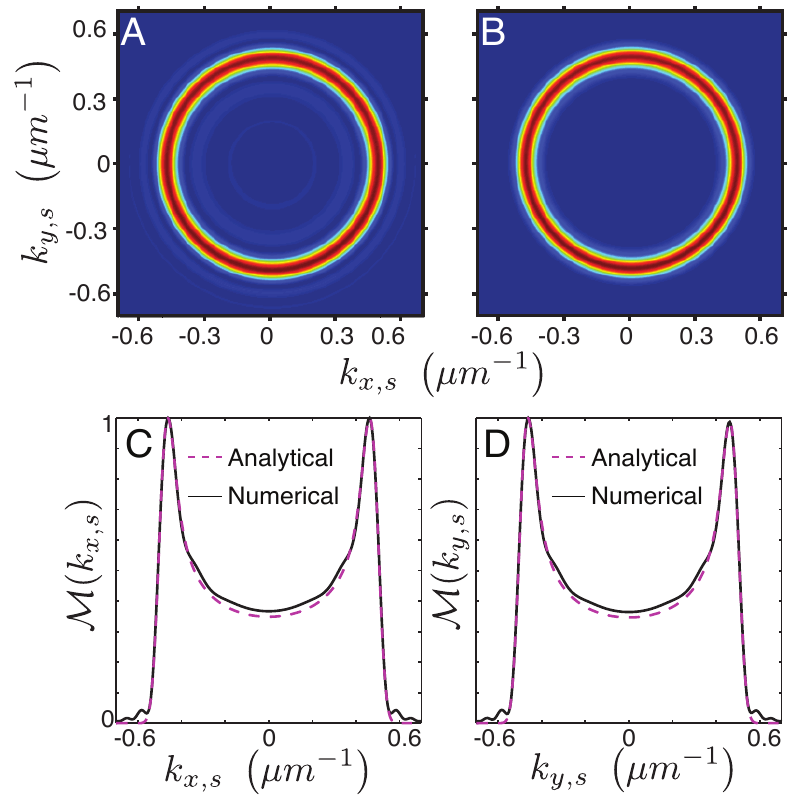}}
\subfloat[]{\label{F:AS35}\includegraphics[width=0.3\textwidth]{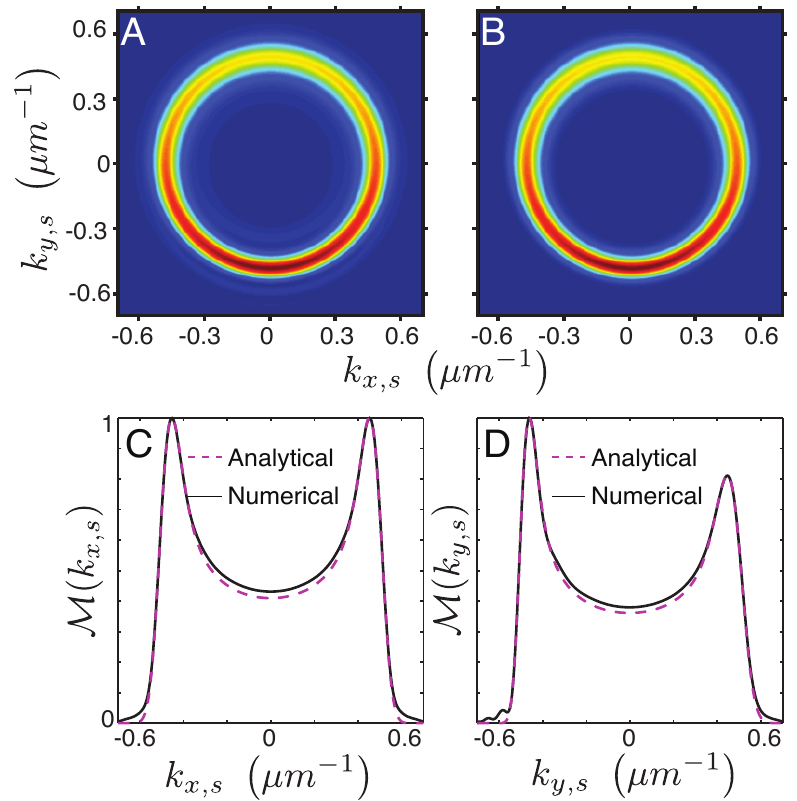}}
\subfloat[]{\label{F:AS5}\includegraphics[width=0.3\textwidth]{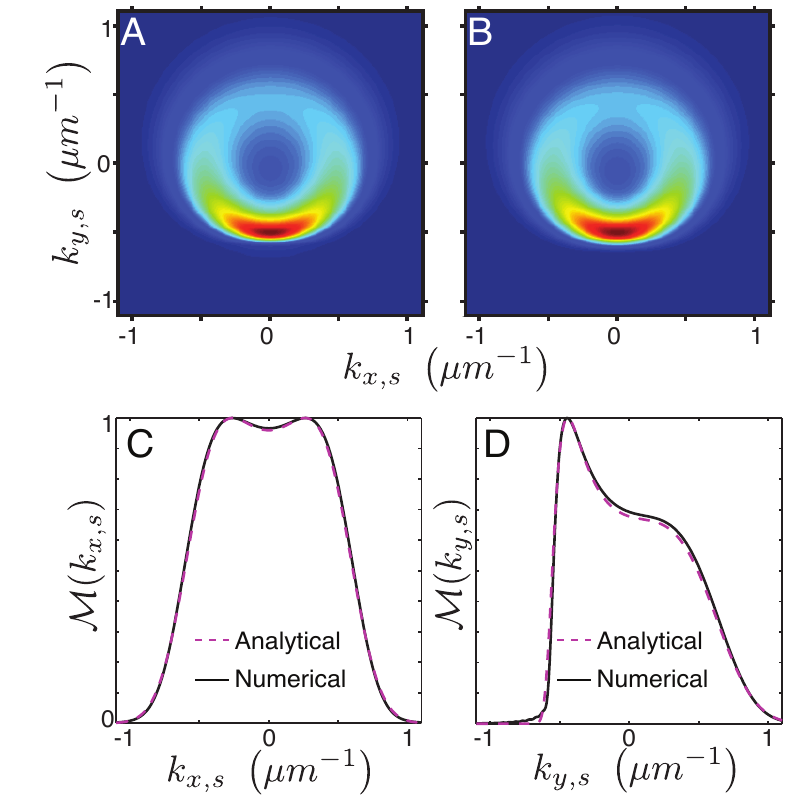}}
\caption{Angular spectrum (AS) for a Gaussian pump with waist (a)$W=185\mu$m; (b)$W=35\mu$m;(c)$W=5\mu$m. In (A) the AS function is calculated numerically and  in (B) using the approximate analytical expression. (C)-(D) are the marginal distributions obtained numerically (solid line) and  analytically (dashed line).}
\end{figure*}

We can now interpret the different terms appearing in Eq.~(\ref{E:RsAn}). For a not too long crystal $u(\tilde{k}_\bot,\theta_{est}) \sim - W^2\tilde k_\bot^2/2$ which does not depend on $k_{\bot,s}$;  in this case $1\gg e^{-u(\tilde{k}_\bot,\theta_{est})}$ just requires $W\gg\tilde k_\bot^{-1}$.
Under these conditions the AS  describes a conical emission with radius $r_{AS}$
 specified just by the difference of the refractive indices. If $n_{eff}(\lambda) =n_o(2\lambda)$, $r_{AS}=0$ corresponding to the collinear regime.  For the particular SPDC source described above  $r_{AS}=0.49 \mu m^{-1}$. The width of the emission  cone $\sigma_{\rm{AS}}$ depends on the value of the pump waist $W$ and the crystal length $L$. In Fig.\ref{F:AS185} (A) we illustrate the AS for the paraxial regime,
$W=185 \mu$m  and $L = 1mm$ and the predicted width of the cone  is $\sigma_{\rm{AS}}=0.03\mu$m$^{-1}$ which coincides, within the given significant figures, with the results obtained numerically using the exact expressions of the dispersion relations and the integrals involved in the calculations.
Due to the Gaussian approximation of the sinc function used to obtain Eq.~(\ref{E:RsAn}) this expression is unable to reproduce faint radial
oscillations of the exact AS. This can be better observed in the marginal distribution with respect the variable $k_{x,s}$ and $k_{y,s}$ illustrated in Figs.\ref{F:AS185} (C)-(D).

We now compare the predictions of Eq.~(\ref{E:RsAn}) for pump beams outside the paraxial regime.
Fig.\ref{F:AS35}(A) illustrates the AS function calculated numerically using a narrower pump waist (
$W=35 \mu$m) and the same crystal parameters described above. The pump beam  focusing  gives rise to an observable  asymmetry in the
cone width that can be attributed to an asymmetric failure of the incorporated wave vectors ${\bf k}_p$ to fulfill the phase matching condition.
 In our approximate expression, this asymmetry is encoded in the factor $L\vert \mathbf{d} \vert /W $ that appears in the expression of $\sigma_{\rm{AS}}$. This factor cannot be neglected out of the paraxial regime even for a narrow crystal ($L$=1mm in Fig.\ref{F:AS35}). Fig.\ref{F:AS35}(B) illustrates the excellent agreement between the numerical and approximate results which in fact also agree with the experimental results reported in Ref.~\cite{uren2013}.
For even higher focused Gaussian beams, the AS function   loses completely its annular symmetry as shown in Fig.~\ref{F:AS5}(A)
where $W=5 \mu m$. The broadening in the $k_{y,s}$ direction is determined by the factor $\beta {\rm a}_y$. Even in
these extreme conditions the approximate  expression Eq.~(\ref{E:RsAn})  reproduces with high accuracy the expectations from the numerical descriptions.

\section{Signal and idler wave vector probability distributions.}

\begin{figure*}
\centering
\includegraphics[width=0.65\textwidth]{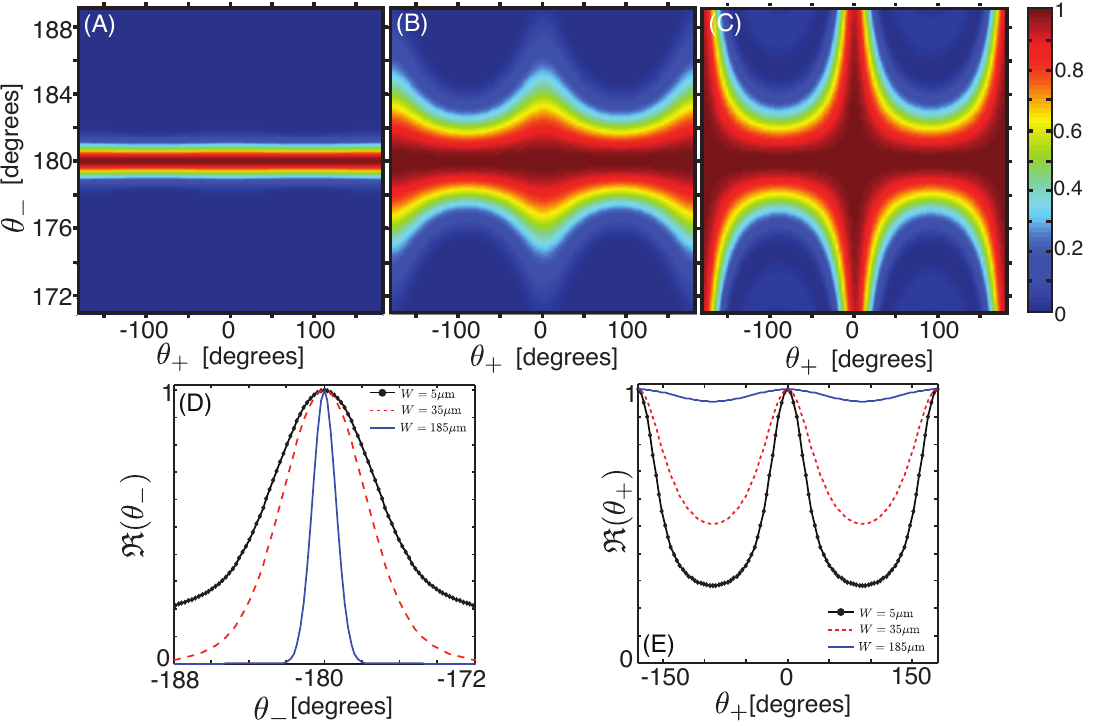}
\caption{Angular correlation function $\mathfrak{R}_{r_{AS}}(\theta_+,\theta_-)$ for a Gaussian pump with waist (A)$W=185\mu$m; (B)$W=35\mu$m;(C)$W=5\mu$m. (D)-(E) are the marginal distributions  normalized to a maximum value one for the same waists of the pump beam. These distributions were obtained using the exact expressions of the dispersion relations and taking the same general parameters as in previous figures.}
\label{F:angcorr}
\end{figure*}

From the approximate analytical expression for the angular spectrum we can  study the marginal distributions on the $ \mathbf{k}_{\bot,s}$ and  $ \mathbf{k}_{\bot,i}$ variables. Since the conditional angular spectrum shows that the idler and signal photons will be emitted nearby a cone with transversal radius $r_{AS}$, defined in Eqs.~(\ref{eq:rassigma}), it results interesting to analyze the relations between the angles of emission $\theta_i$ and $\theta_s$ of the idler and signal photons in that cone. We find easier to make such an study in terms of the mean angle, $\theta_+ = (\theta_i +\theta_s)/2$  and the difference angle $\theta_-=
\theta_s - \theta_i$. The distribution
\begin{equation}
\mathfrak{R}_{r_{AS}}(\theta_+,\theta_-) = |F(\mathbf{k}_{\bot,s},\omega_s ;\mathbf{k}_{\bot,i},\omega_i)|^2/\vert g L \alpha(\omega_p) \vert^2,
\end{equation}
$$\mathbf{k}_{\bot,s} =r_{AS}(\cos(\theta_+ +\theta_-/2),\sin(\theta_++\theta_-/2)),$$
$$\mathbf{k}_{\bot,i} =r_{AS}(\cos(\theta_+ -\theta_-/2),\sin(\theta_+-\theta_-/2)),$$
is illustrated for the pump beam waists $W=185 \mu$m, $W=35\mu$m and $W=5\mu$m in Fig.\ref{F:angcorr}, taking the same general parameters used in the simulations reported before; in particular the crystal cut angle is $\theta_{\mathbf{a}}=29.3^\circ$
yielding $r_{AS}^{29.3^\circ} = 0.492\mu$m$^{-1}$. The exact dispersion relations for the pump, idler and signal photons were used. The calculation results coincide with those obtained using the approximate dispersion relations for $W>25\mu$m within the  first three significant figures. Notice that the idler and signal photons are emitted mainly in opposite transverse directions since the maximum value of  $\mathfrak{R}_{r_{AS}}(\theta_+,\theta_-) \sim 1$ is always taken for
$\theta_- = 180^\circ$. However, for $W<25\mu$m the width of the marginal distribution of $\theta_-$ increases as the waist decreases, so that there is a high possibility that the signal and idler photons are emitted with $\theta_-$  $\in$ $(175^\circ,185^\circ)$. It can also be observed that the emission cone will have an almost isotropic distribution of photons for $W>100\mu$m since for those values of the pump waist  $\mathfrak{R}_{r_{AS}}$ shows only a slight dependence  on $\theta_+$. For lower values of $W$, the largest values of the marginal distribution of $\theta_+$ are achieved for  $\theta_+ = 0,\pm\pi$. If $\theta_-\sim\pi$, this means that $\theta_{i,s}\sim \pm\pi/2$, which  is consistent with all the results for the CAS shown in last section.

The approximate analytical expression for the CAS distribution, Eq.~(\ref{E:RsAn}), can be used to obtain approximate expressions for the mean aperture angle  of the emission cone $$\langle\Theta_{AS}\rangle =\sin^{-1}\Big(r_{AS}/\sqrt{(\omega n_o/2c)^2 -r_{AS}^2}\Big).$$
The spread in the distribution of this angle is determined by the spread in the distribution of the cone radius  $\sigma^{1/2}/\sqrt{2}$, given in Eq.~(\ref{eq:rassigma}), and it depends on the direction of the wave vectors $\mathbf{k}_{s,i}$.
Its maximum (minimum) value is achieved for signal photons emitted at angles that maximize (minimize) the magnitude of the $\mathbf{d}$ vector. For a signal photon  with $\vert\mathbf{k}_s\vert \sim r_{AS}$, these extreme values are given by $\vert \mathbf{d}_{ext}\vert\sim\vert\beta\cos\alpha \mp(2c/n_o\omega)r_{AS}\vert$. The corresponding extreme values of the aperture angle are
$$\Theta^{ext}_{AS} = \langle\Theta_{AS}\rangle + \Delta\Theta_{AS}^{ext}$$
\begin{equation}
\Delta\Theta_{AS}^{ext} \sim \frac{(1+(\gamma L\vert \mathbf{d}_{ext}\vert/W)^2)^{1/4}}{2^{5/4}\sqrt{\gamma L(\omega n_o/2c)}}.
\end{equation}
For the general parameters used up to now, $\langle\Theta_{AS}\rangle= 0.038$ rad and
 for $W=185\mu$m the spread $\Delta\Theta^{max}_{AS} = 0.006$ rad $\sim\Delta\Theta^{min}_{AS} $,  for $W=35\mu$m the maximum spread increases to $\Theta^{max}_{AS} = 0.009$rad and the minimum $\Theta^{min}_{AS} = 0.007$ rad, and for the extreme condition $W=5\mu$m, $\Delta\Theta^{max}_{AS} = 0.022$ rad and $\Delta\Theta^{min}_{AS} = 0.017$ rad.

\section{SPDC optimization.} 

It has been shown that SPDC with extremely focused Gaussian beams in general gives rise to an anisotropic distribution
of photon pairs in an emission cone with a mean radius independent of the pump waist, but with a spread that may
increase as the pump waist diminishes. This fact may make difficult the separation of the idler and signal effective modes.
The SPDC simulations reported in  previous sections were worked out with parameters chosen to optimize a type I 
SPDC process for an ideally incoming plane wave. 
In this section we show that parameters like cut angle of the nonlinear crystal or the wavelength of the pump beam can be easily optimized to increase the emission cone radius,
preserve approximately the spread of the cone radius and simultaneously preserve the anisotropic localization of  the idler and signal wave vectors  without a significant loss in the probability of coincidence detection of the emitted photons.
 
In standard experimental set--ups, the crystal cut angle is chosen so that an ideal pump beam with an almost plane wave front
yields the possibility of creating almost collinear photon pairs with maximum probability. By increasing the cut angle, the effective extraordinary refraction index $n_{eff}$
decreases and the mean radius of the emission cone $r_{AS}$ increases. Thus, it is reasonable to study the behavior of SPDC for higher values of  $\theta_{\mathbf{a}}$ to decrease the overlap between the idler and signal effective modes. Notice however that $\mathbf{d}_{ext}$ also depends on  $\theta_{\mathbf{a}}$, and we 
are interested in finding optimal values of $\theta_{\mathbf{a}}$ that do not increase the width of the emission cone beyond the adequate values for the apertures of usual detecting elements. In most cases, the latter  include optical fibers to which the signal and idler photons are required to couple.

\begin{figure*}
\centering
\includegraphics[width=0.7\textwidth]{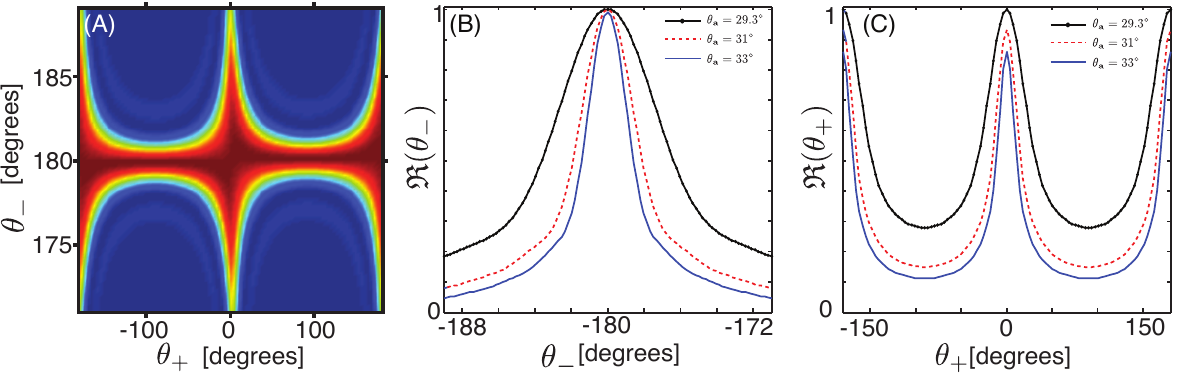}
\caption{Angular correlation function $\mathfrak{R}_{r_{AS}}(\theta_+,\theta_-)$ and marginal distributions on the relative angle $\theta_-$ and average angle $\theta_+$ for a Gaussian pump waist $W=5\mu$m
and different cut angles $\theta_{\mathbf{a}} = 29.3^\circ$, $\theta_{\mathbf{a}} = 31.0^\circ$ and $\theta_{\mathbf{a}} = 33.0^\circ$. These distributions were obtained using the exact expressions of the dispersion relations and taking the same general parameters as in previous figures.}
\label{F:andop1}
\end{figure*}

Using the  formalism presented in last section, the SPDC process were roughly simulated and optimal angles were easily found.
In Fig.~(\ref{F:andop1}) we report the behavior of the  $\mathfrak{R}_{r_{AS}}(\theta_+,\theta_-)$ and its marginals
for a pump beam with $W=5\mu$m for $\theta_{\mathbf{a}}=29.3^\circ, 31.0^\circ$ and $33.0^\circ$. The 
mean emission cone radii are $r_{AS}^{31.0^\circ} = 0.955\mu$m$^{-1}$ and $r_{AS}^{33.0^\circ} = 1.313\mu$m$^{-1}$, so that $\langle\Theta_{AS}^{31.0^\circ}\rangle = 0.075 $rad  and $\langle\Theta_{AS}^{33.0^\circ}\rangle
= 0.103$ rad. The angular maximum and minimum widths are $\Delta\Theta^{max,31.0^\circ}_{AS} = 0.024$ rad and $\Delta\Theta^{min,31.0^\circ}_{AS} = 0.013$ rad, and $\Delta\Theta^{max,33.0^\circ}_{AS} = 0.026$ rad and $\Delta\Theta^{min,33.0^\circ}_{AS} = 0.009$ rad. So that, increasing slightly the cut angle $\langle\Theta_{AS}\rangle$ can take two or three times its original value while the maximum value of $\Delta\Theta_{AS}$ is modified in less than 15$\%$.
Notice that the change in the maximum value of the marginal distributions evaluated at $r_{AS}$ change in less than $15\%$.

\begin{figure*}
\centering
\includegraphics[width=0.7\textwidth]{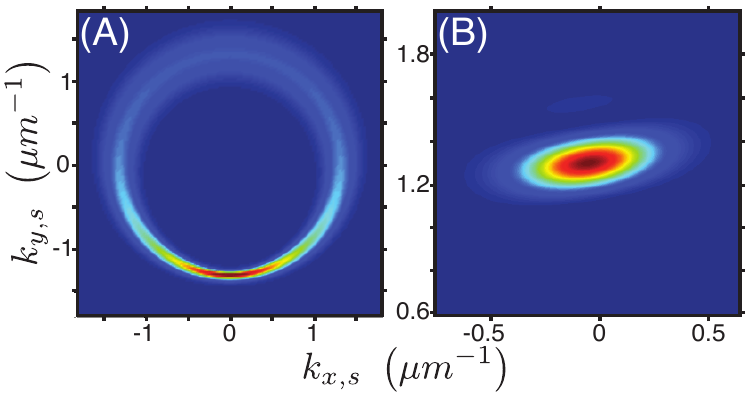}
\caption{(A)AS and (B) CAS for a Gaussian beam with waist $W=5\mu$m and a cut angle of the nonlinear crystal $\theta_{\mathbf{a}} = 33^\circ$ optimized. These distributions were obtained using the exact expressions of the dispersion relations and taking the same general parameters as in previous figures.}
\label{F:andop2}
\end{figure*}

Taking into account those results we have performed a complete simulation of the SPDC process for the pump beam with
$W=5\mu$m and a cut angle $\theta_{\mathbf{a}}=33^\circ$. The corresponding AS and CAS are shown in Fig.~(\ref{F:andop2}).
As predicted from the simulations described in the last paragraph, the spatial resolution of the idler and signal photons has greatly increased while the location where the emitted photons is still concentrated around $\theta_{s,i}\pm\pi/2$.
The numerical evaluation of the brightness calculated from the CAS function shows that it increases in a $10\%$ for the cut angle   $\theta_{\mathbf{a}}=33^\circ$ compared to  $\theta_{\mathbf{a}}=29.3^\circ$. Besides the rough estimates for the
cone angle and its width given in last paragraph agree approximately with the exact numerical calculation.

We have also studied the possibility of maintaining the original cut angle $\theta_{\mathbf{a}}=29.3^\circ$ and optimize the pump wavelength for highly focused Gaussian beams. The first calculation using the expression for $r_{AS}$ and $\mathbf{d}$ give us that
changing the pump wavelength from $406.99\mu$m to $436\mu$m  ($483\mu$m) yields the same results than changing to $\theta_{\mathbf{a}}=31^\circ$ ($\theta_{\mathbf{a}}=33^\circ$).

\section{Conclusions}

In this paper we have presented a study of the dependence of the angular spectrum and the conditional angular spectrum on the waist  pump and the length of a nonlinear crystal used to achieve type I SPDC. Simple analytic expressions were  derived for the these spectra, Eqs.~(\ref{E:F2}-\ref{eq:rassigma}). They make evident  the relevance of  the factor $\gamma\vert{\mathbf d}\vert L/W$,  which includes information about the walk off, to understand the
general features of these spectra. Notice that under the conditions here studied the CAS and the AS are independent on the position of the focus plane
of the pump beam. However it could lead to observable effects in other properties of the down converted photons like the purity of state. The analytical expressions  reproduce with high accuracy the predictions of numerical methods that do not make use of the approximations neither on the dispersions relations nor on the phase matching function. Even more important, the results presented here are in excellent agreement with  experimental measurements reported in the literature.

Having an analytical expression for the CAS allows the identification of explicit expressions for the mean traverse radii  and the anisotropic width of the emission cone in the wave vector space. The first depends just on the linear dielectric response and the cut angle of the birefringent crystal. The anisotropy depends on the relation between the crystal length and beam waist
through the factor $\gamma\vert{\mathbf d}\vert L/W$. Given a crystal length, a highly anisotropic SPDC process for extreme focused Gaussian pump beams is especially relevant since it allows the prediction of the direction at which idler and signal photons will be generated. This anisotropy has been already reported by Fedorov $et$ $al$ in Ref.~\cite{Fedorov2007,Fedorov2008}, where special care
is taken of the theoretical and experimental analysis of the CAS and the AS along  two lines one of which is contained in a plane parallel to the optical axis and the other in a plane perpendicular to that axis. In Ref.~(\cite{uren2013}),  the anisotropy is also reported in terms of the AS and the CAS, so that the comparison between their experimental and our theoretical results could be directly performed.

We have used also our expressions to make estimates of the general properties of the distribution of idler and signal photons in wave vector space. They can be used to measure entanglement \cite{Fedorov2007,Fedorov2008} and for an optimization of the SPDC process. That is, by simple calculations that nevertheless include the exact expressions for $\Delta k_z=k_{z,p}-k_{z,s}-k_{z,i}$ evaluated on $r_{AS}$ the angular distribution of the emitted photons can be studied and conditions for a given angular localization can be found; using the expression of $\sigma_{AS}$ the maximum and minimum spread the idler and signal photons may be studied looking for values that could, $e.$ $g.$ optimize the coupling to optical fibers without a significant loss in the probability of coincidence detection of the emitted photons. 
Our analysis complements the studies reported in Refs.~\cite{Bovino2003, dragan2004,Castelletto2005,Andrews2004,Ljunggren2005,Kolenderski2009, minozzi2013}, providing a simple way to preselect the general experimental parameters in an optimization procedure. These parameters can be then used to make a detailed simulation of the process, previous to the experimental implementation.

\end{document}